# Further evaluation of a THGEM UV-photon detector for RICH – comparison with MWPC

**V. Peskov**[a,b*]**, M. Cortesi**[a]**, R. Chechik**[a] **and A. Breskin**[a]

[a] *Department of Particle Physics, Weizmann Institute of Science,*
  *76100 Rehovot, Israel*
[b] *CERN,*
  *1206 Geneva, Switzerland*

  *\*corresponding author*
  *E-mail*: Vladimir.Peskov@cern.ch

ABSTRACT: The operation of single-, double- and triple-THGEM UV-detectors with reflective CsI photocathodes (CsI-THGEM) in Ne/$CH_4$ and Ne/$CF_4$ mixtures was investigated in view of their potential applications in RICH. The studies were carried out with UV, x-rays and β-electrons and focused on the maximum achievable gain, discharge probability, cathode excitation effects and long-term gain stability. Comparative studies under similar conditions were made in $CH_4$, $CF_4$ and Ne/$CF_4$, with a MWPC coupled to a reflective CsI photocathode (CsI-MWPC). It was found that at counting rates $\leq 10$ Hz/mm$^2$ the maximum achievable gain of CsI-THGEMs is determined by the Raether limit; at counting rates $> 10$ Hz/mm$^2$ it dropped with rate. In all cases investigated the attainable CsI-THGEM gain was significantly higher than that of the CsI-MWPC, under similar conditions. Furthermore, the CsI-THGEM UV-detector suffered fewer cathode-excitation induced effects as compared to CsI-MWPC and had better stability at high counting rates.

# Contents

## 1. Introduction

The present study aimed at evaluating some properties of CsI-coated Thick Gaseous Electron Multipliers (THGEM [1]) as robust UV-photon detectors for Ring Imaging Cherenkov (RICH) applications [2]. There are several advantages for that:

1. High gains ($> 10^5$) are reachable with single- or cascaded-THGEM electrodes in selected gases; mixtures of choice could be Ne with a small addition of quenchers ($CH_4$, $CF_4$) [3, 4]. Due to the exponential nature of single-photoelectron pulse-height distributions, and taking into account signal-over-threshold considerations, a high detector gain is an important factor in improving single-photon detection efficiency.

2. A THGEM can operate in poorly quenched gas mixtures as well as in gases emitting UV light (e.g. noble gases [5], $CF_4$ [6, 4]). This permits conceiving windowless detectors (same detector and radiator gas, e.g. like in [7]), with simpler layout and larger Cherenkov-photon detection yields.

3. In intense-background environment, THGEMs can operate in the so-called "Hadron-Blind mode" with zero or reversed electric field above the photocathode [8]; this significantly reduces particle-induced ionization signals [7].

CsI-THGEMs are currently considered for the upgrade of the RICH systems at the CERN-ALICE and -COMPASS experiments [9, 10]. The reasons to replace the CsI-MWPCs [11] in these experiments are numerous: The CsI-MWPCs operate at relatively low gas gains ($\sim 10^4$), limited by avalanche photon-feedback. Furthermore, there is evidence that at high counting rates the MWPCs suffer from feedback-related discharges, followed by difficulties to restore the operating voltage, up to 1-day periods [12]. This could have resulted from a cathode excitation process [13], in which the surface is modified under discharge-induced ion bombardment, with an increase of the quantum efficiency of the cathode; this can trigger a feedback loop, followed by discharges and further by an on-going electron-jet emission from the affected cathode region. Ion-induced photocathode degradation in the open-geometry of wire chambers, affecting the detector lifetime, would be strongly reduced in CsI-THGEMs, where only a small fraction of avalanche ions hit the photocathode. However, replacing the CsI-MWPC currently used in these and in other experiments [11] should be carefully assessed and validated.

We have recently initiated a series of measurements [3, 4], systematically studying the gain and single-photoelectron detection efficiency of CsI-THGEMs, in view of their potential application in UV detectors for RICH systems [6]. We have identified two gas mixtures, Ne/CH$_4$ and Ne/CF$_4$, adequate for stable operation of CsI-THGEM; we demonstrated that in these mixtures photoelectron extraction efficiencies from the reflective CsI photocathode reach about $\varepsilon_{extr}$ ~ 80% under proper HV values across the holes; all the extracted photoelectrons were collected into the THGEM holes, $\varepsilon_{coll}$ ~ 100%, where they experienced gas multiplication [4]. The effective photon detection efficiency ($E_{effph}$) of the photosensitive THGEMs is:

$$E_{effph}(\lambda) = QE(\lambda)\ A_{eff}\ \varepsilon_{extr}(\lambda)\ \varepsilon_{coll} \qquad (1)$$

where $QE(\lambda)$ is the CsI quantum efficiency in vacuum as a function of the wavelength $\lambda$, $A_{eff}$ is the CsI-coated fraction of the THGEM surface. As was shown in [4], $E_{effph}$ could reach values (at $\lambda = 170$ nm, with $QE = 30\%$) in the range of 0.13-0.24 (depending on $A_{eff}$). The expected photon detection efficiency of CsI-THGEM could therefore reach values of $E_{eff} \approx 0.21$ (assuming 90% photoelectron detection efficiency above threshold, due to the high gain); it is to be compared to that of the CsI-MWPC presently employed in COMPASS-RICH ($E_{eff} \approx 0.21$ with $QE = 30\%$ and 70% photoelectron efficiency [12]).

In the present work we further elaborate on the maximum achievable gain before the onset of feedback or discharges, particularly, detecting UV photons in the presence of ionizing radiation (X-rays or β-electrons). Comparative results with CsI-MWPCs are provided, including the operation of MWPCs and THGEMs in CF$_4$ in view of their application in a windowless UV-detector for RICH.

## 2. Instabilities in gas-avalanche detectors

As follows from several previous investigations (see [13] and references therein) the breakdown in micropattern gas detectors (uncoated with CsI), including THGEMs, is determined by the Raether limit [14]; it occurs when the total charge in the avalanche exceeds a certain limit:

$$G_m\ n_0 = Q_{crit} \qquad (2)$$

where $G_m$ is the maximum achievable gas gain, $n_0$ is the number of primary electrons created by the ionizing radiation. Depending on the detector type and the gas mixture $Q_{crit} = 10^6 - 10^7$ electrons.

Equation (2) stresses the fact that indeed very high gains, e.g. $G_{ph}$ ~ $10^6$-$10^7$, may be obtained with single UV photons, under very low ionizing radiation background. But under intense background, the gain may drop by as much as two orders of magnitude [3]. This effect is of course expected to be more pronounced in "heavier" gas mixtures, e.g. Xe, Kr and Ar, than in Ne-mixtures having lower particle-induced ionization-electron yields [13]. In addition, $G_m$ may also drop with increasing counting rate [13]. To our best knowledge, systematic studies of these effects in cascaded photosensitive THGEMs were not done so far.

In gaseous detectors coupled to efficient photocathodes, photon- and ion-induced positive feedback may also trigger breakdown. This happens, when

$$G_f \, \gamma_f = 1 \qquad (3)$$

where $G_f$ is the maximum achievable gain in these conditions and $\gamma_f$ is the probability of a secondary-electron creation by the feedback mechanism (we specify $\gamma_{fp}$ and $\gamma_{fi}$ in relation to photon- and ion-induced electron emission, and $G_f$ and $G_i$ as maximum achievable gain respectively). Measurements with various open-geometry photosensitive detectors (wire-chambers or parallel-plate ones) showed that usually $G_f < G_m$ [13]. In contrast, in hole-type multipliers, the majority of the avalanche photons are masked by the hole walls, and only a small fraction $k_{fp}$ ($k_{fp} \ll 1$) impinges on the photocathode; the condition for photon-feedback breakdown is:

$$G_f \, k_{fp} \, \gamma_{fp} = 1 \qquad (4)$$

Ion-backflow is also reduced in hole-type multipliers, especially in cascaded THGEMs, and only a fraction $k_{fi} < 1$ of ions reaches the photocathode. Thus the condition for ion-feedback breakdown will be

$$G_i \, k_{fi} \, \gamma_{fi} = 1 \qquad (5)$$

Thus feedback-induced breakdown is significant when $\gamma_{fi}$ is large, e.g. with photocathodes sensitive in the visible-light range [15, 16], having high values of $\gamma_{fp}$ and $\gamma_{fi}$.

One of the goals of this work was to assess the probability for photon-induced and ion-induced breakdown in cascaded THGEMs with reflective CsI photocathodes.

Another source of gain instabilities investigated here is the cathode excitation effect; it is related to the feedback loop [13] and prevents the immediate restoration of the detector's high voltage following a discharge.

## 3. Experimental setup

A schematic drawing of our experimental setup is shown in Fig. 1. It consists of a vacuum chamber housing a triple-THGEM detector, a MWPC detector included for comparative purposes, a gas system allowing gas circulation with various gas mixtures, a high vacuum (typically $10^{-5}$ Torr) pumping system, a UV lamp, a lamp emitting only in the visible range, an X-ray tube and several radioactive sources. The chamber has appropriate windows on both sides, to allow operation of each detector separately with any of the above radiation sources.

The triple-THGEM layout, which is similar for the bare and the CsI-coated one, is shown in Fig. 2. The electrodes had the following geometry: thickness $t = 0.4$ mm, hole-diameter $d = 0.3$ mm (or 0.4 mm in some measurements), holes rim $h = 0.1$ mm and holes-pitch $a = 0.8$ mm (for $d = 0.3$ mm) and $a = 1$ mm (for $d = 0.4$ mm). The active areas of the THGEMs investigated were: $2.5 \times 2.5 \text{ cm}^2$ and $10 \times 10 \text{ cm}^2$. The conversion gap above the first THGEM was of 10 mm, the transfer and induction gaps were 2 mm wide. The electrodes were individually biased by CAEN N471A power supplies. Measurements were performed both in current and in pulse-counting modes; currents were recorded with sensitive picoampermeters (Keithley 610CR), pulses were processed with a charge-sensitive preamplifier Ortec PC140 and with a fast current amplifier VT110CH4 (ESN Electronics). In most of the measurements the top electrode of the first THGEM was coated with a 300 nm thick reflective CsI photocathode, vacuum deposited onto its Au-coated surface.

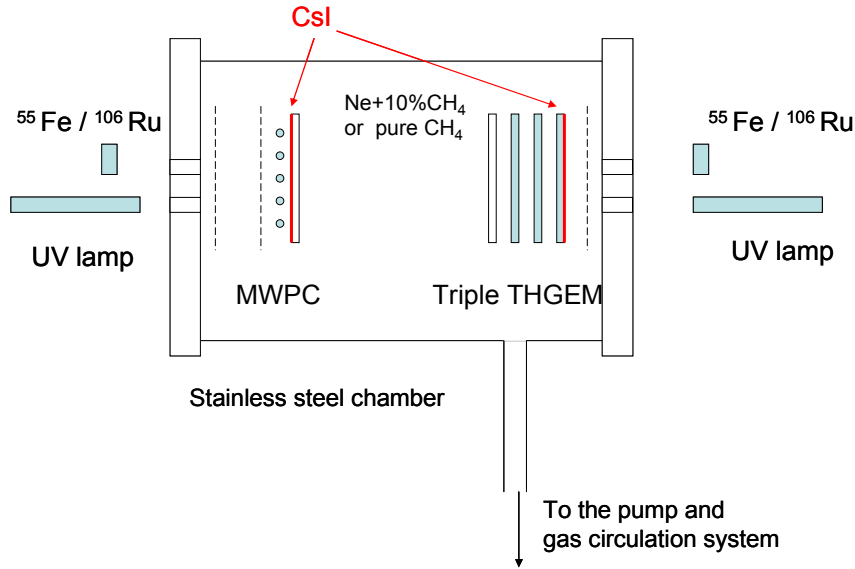

**Figure 1.** A schematic drawing of the experimental setup, enabling the comparative investigation of CsI-MWPC and CsI-THGEM with single photons, x-rays and β- electrons.

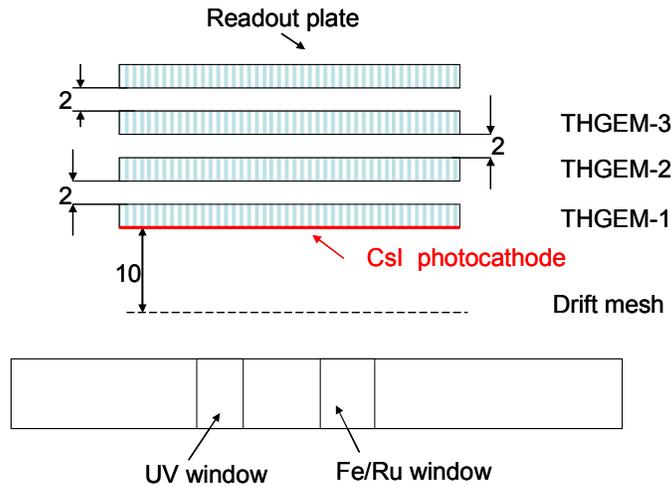

**Figure 2.** Schematics of the triple THGEM detector used in this work. See text for details of the THGEM electrode's geometry.

The detectors were operated under continuous gas flow, at 1 atmosphere; Ne/CH$_4$ and Ne/CF$_4$, with ratios of (95:5) and (90:10) were used. The gases were of 99.995% purity. The gas composition and its flow were controlled with Mass Flow Controllers (MKS type 1179A) and a control/readout module (MKS type 247). Some measurements were performed in pure CF$_4$ at 1 atm.

The layout of the 2.5×2.5 cm$^2$ MWPC detector is shown in Fig. 3. It had 20 μm in diameter anode wires, at 3 mm pitch; 50 μm in diameter wires were placed, at 1 mm pitch, at the edges of the anode frame. The measurements were performed both with bare and with 300 nm thick (vacuum deposited) CsI-coated stainless-steel (SS) cathode. The UV source was a Ar(Hg) lamp (Oriel); the visible-light source was a small, battery-powered torch.

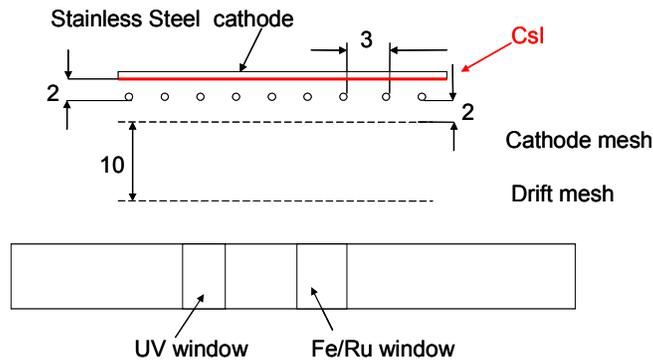

**Figure 3.** A schematic drawing of the MWPC used for comparative studies.

Two x-ray guns were used at various stages of these studies: a micro-focus (35 μm) x-ray tube (XTF5011 - Oxford Instruments plc) having a Cu anode and a Kevex x-ray tube with a Mo anode. The tubes produced Bremsstrahlung photons, of broad energy spectra, with peaks at 9 and 17.5 keV respectively. The beam geometries were defined by cylindrical apertures placed in front of their anodes, with openings of 2-4 mm in diameter. The radioactive sources used in some measurements were $^{55}$Fe, $^{106}$Ru, $^{90}$Sr and $^{241}$Am.

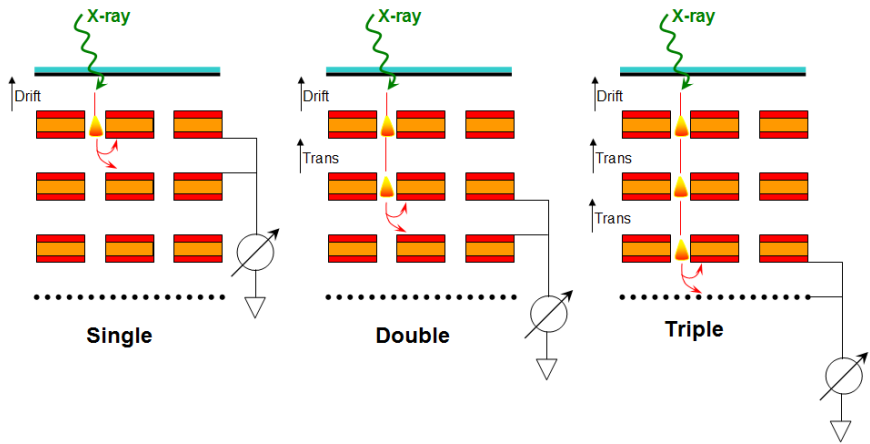

**Figure 4**. Schematic layouts of the single-, double- and triple-THGEM configurations used for gain measurements as function of radiation (X-rays) flux.

The gas gain measurements in single- and cascaded-THGEM detectors were performed in current mode, recording currents from interconnected electrodes in each of the configurations shown in Fig. 4, with a Keithley 610 CR electrometer.

## 4. Results
### 4.1 Results with THGEMs
#### 4.1.1 Bare THGEMs

In order to monitor the gain variation as function of the x-ray flux rate we have first measured the gas gain under x-ray irradiation as function of the voltage applied across the THGEMs, in the 4 different gas mixtures, in single-, double- and triple-THGEM detectors (Fig. 5). The drift field was set at 0.6 kV/cm; for double- and triple-THGEMs the transfer electric field was 0.5 kV/cm. The gain curves were measured under relatively low x-ray flux, of 100 Hz/mm$^2$.

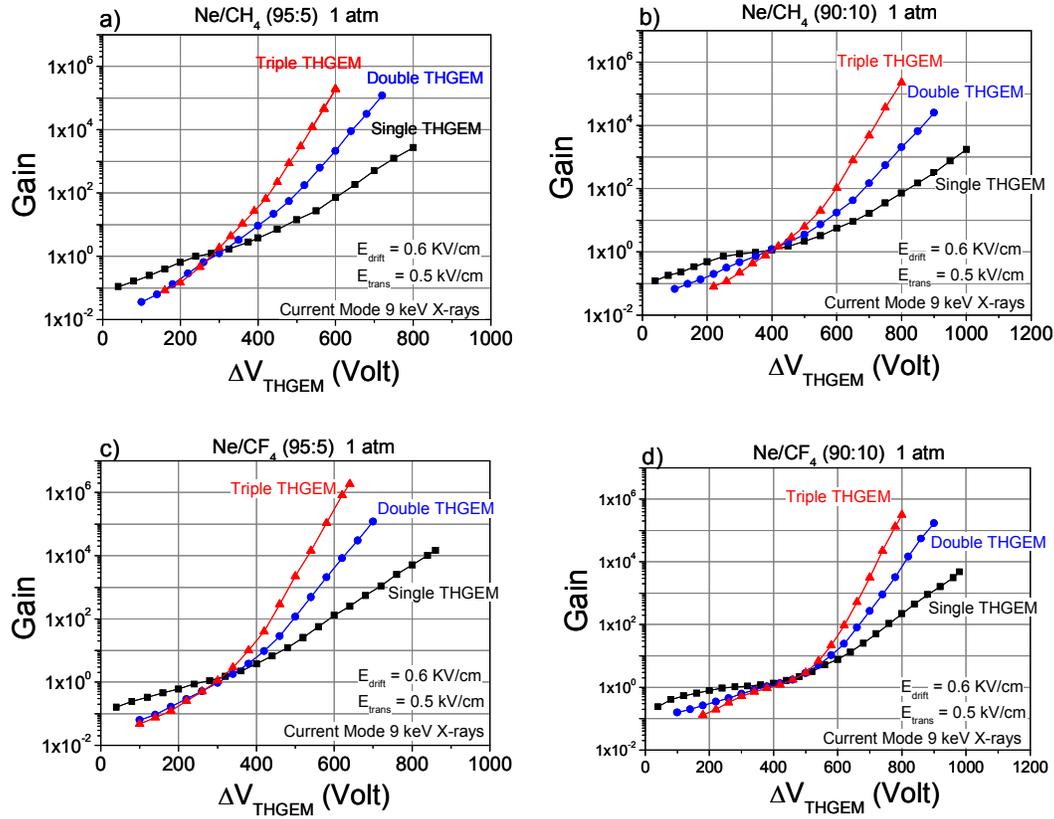

**Figure 5.** Gain-curves measured with low flux (~100 Hz/mm$^2$) of 9 keV x-rays, in the single-, double- and triple-THGEM detectors of Fig. 4, having hole-diameters of 0.3 mm, 0.8 mm pitch, 0.1 mm rim and 0.4 mm thickness, operating at atmospheric pressure in: a) Ne/CH$_4$ (95:5); b) Ne/CH$_4$ (90:10); c) Ne/CF$_4$ (95:5) and d) Ne/CF$_4$ (90:10).

As already observed, the operation voltages and the maximum achievable gains in Ne/CH$_4$ and in Ne/CF$_4$ were rather similar [3, 4], with increasing operation voltages at higher quencher quantities; e.g. in single-THGEM the maximum gain (few times $10^3$) was reached at ~ 800 V with 5% of quencher and at ~ 1000 V with 10%. As observed in previous works on cascaded GEMs or THGEMs, the larger was the number of multipliers in the cascade the lower was the operational voltage of each individual multiplier - resulting in better stability and higher reachable gains. It has also been confirmed that due to electron-diffusion effects [13] the Raether limit in cascaded detectors is generally higher.

Figure 6 depicts the gain drop as function of the counting-rate for single-, double- and triple-THGEM configurations, in 5% and 10% mixtures of Ne/CH$_4$ (and Ne/CF$_4$). The gain drop may be due to either accumulation of space-charge or to charging-up processes, both depending strongly on the gas gain. The maximum gains at which discharges appeared (indicated by arrows in Fig. 6) follow the general trend empirically established in other gaseous detectors [13]; the region above this limit (spark region) is dominated by continuous gas breakdowns. The arrows in the figure refer to x-ray-induced discharge onset.

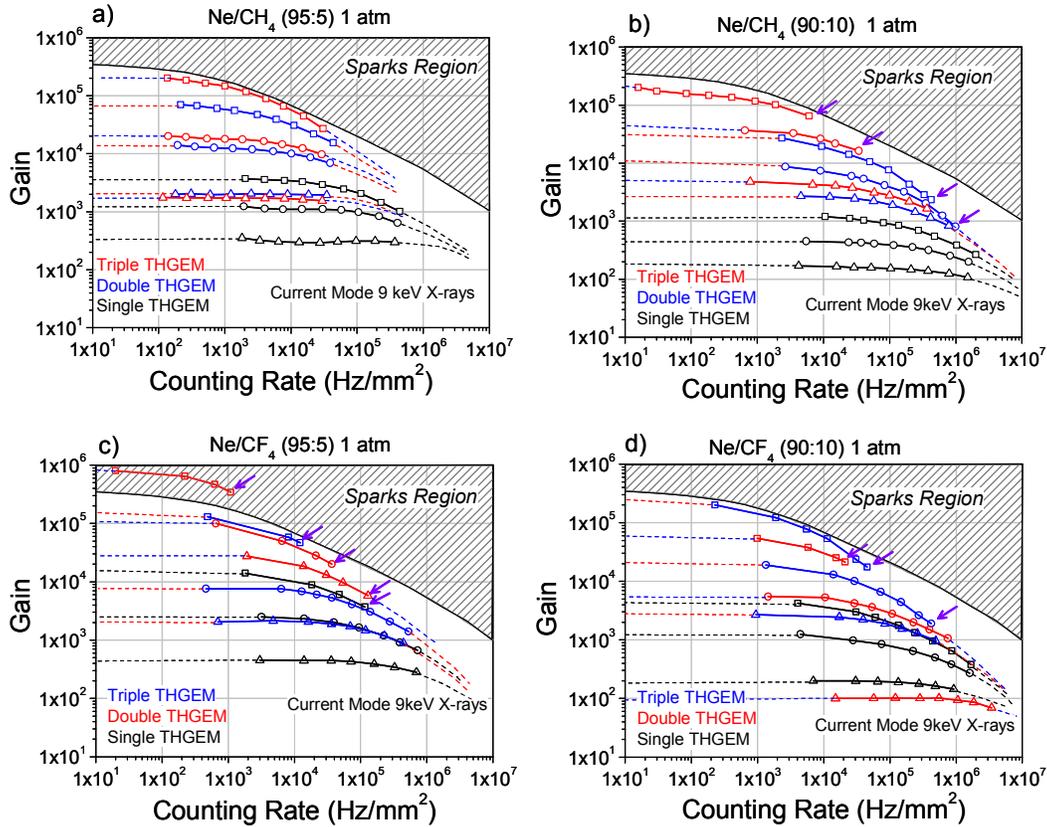

**Figure 6.** Dependence of the gain, measured with 9 keV X-rays, on counting rate, for single-, double- and triple-THGEM detectors (of Fig. 4) operating in: a) Ne/CH$_4$ (95:5); b) Ne/CH$_4$ (90:10); c) Ne/CF$_4$ (95:5) and d) Ne/CF$_4$ (90:10). The arrows indicate the onset of breakdown.

The graphs in Fig. 6 indicate that, in the present quencher-concentration range, for any given gain value, the three THGEM configurations behave quite similarly; the curves tend to overlap at gain × rate values exceeding $10^9$. Note that in Ne/5%CH$_4$ the increase of rate did not induce breakdowns. In our voltage setting scheme, namely equal voltages on all THGEM electrodes of a cascade, the breakdowns were always observed in the last multiplier in the cascade.

### 4.1.2 CsI-THGEMs

*Maximum gain and breakdown rates at low-intensity background.*

Figure 7 shows gain curves, measured in current mode with UV photons, in single-, double- and triple-THGEMs, irradiated simultaneously with the UV lamp and with either $^{55}$Fe or $^{106}$Ru. The respective counting rates over the ~ 1 cm$^2$ irradiated area were: $N_{UV}$ ~ 10$^5$ Hz, $N_{Fe}$ ~ 10$^2$-10$^3$ Hz and $N^{Ru}$ ~ 30 Hz. The highest point on each gain curve corresponds to the onset of discharges (~ 1 per 10 min).

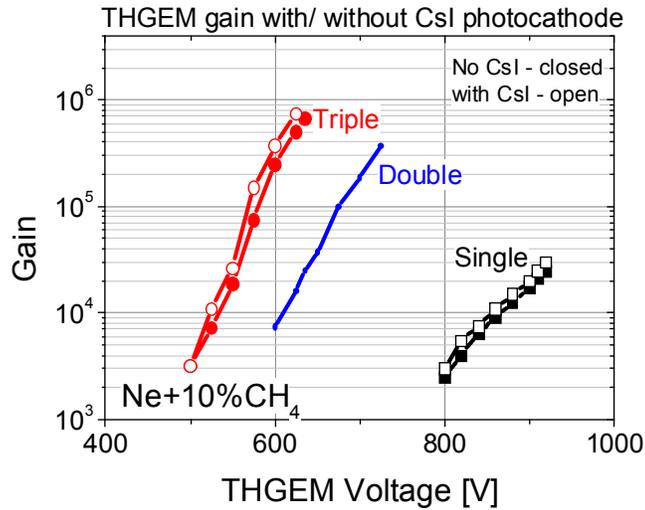

**Figure 7.** Gain curves measured with single-, double- and triple-THGEMs having hole-diameter of 0.4mm, in Ne+10%CH$_4$. The detectors were irradiated simultaneously with UV light and with a radioactive source, of either $^{55}$Fe or $^{106}$Ru. Results are shown for detectors with CsI photocathode (open symbols) and without CsI (filled symbols).

These measurements are very relevant for photon detectors operating under intense background of highly ionizing radiation; the higher-ionization background events reaching the Raether limit, are responsible for the discharge onset. As reflected from Fig. 7, similar results (slope, maximum gain) were reached in measurements done with bare and with CsI-coated THGEMs. This indicates that the CsI coating does not induce noticeable photon feedback (see discussion), thus the maximum achievable gain being determined only by the Raether limit, as in bare THGEMs

In a next set of measurements the discharge-rate as function of the voltage/gain was investigated in two configurations: normal and reversed drift fields above the photocathode. The reversed drift field configuration has been proposed (see [8] and references therein) as means for reducing the detector's sensitivity to ionization electrons deposited in the gas within the drift gap, mostly by background radiation, e.g. Hadronic background in Cherenkov detectors. The effectiveness of this approach was demonstrated and confirmed in beam operation, e.g. with the cascaded-GEM photon detectors of the Hadron-Blind (Cherenkov) Detector (HBD) of the PHENIX experiment [7], where the background sensitivity was shown to be reduced ~ 10 fold.

The present measurements focused on the effect of the reversed-field configuration in reducing the background-induced discharge rate. The detector was illuminated with UV photons and with UV plus beta-electrons or 6 keV x-rays. Because of the sharp drop of the discharge rate with decreasing gain, the counting statistics in these laboratory measurements was rather poor; nevertheless, the general trend can be clearly observed in Fig. 8.

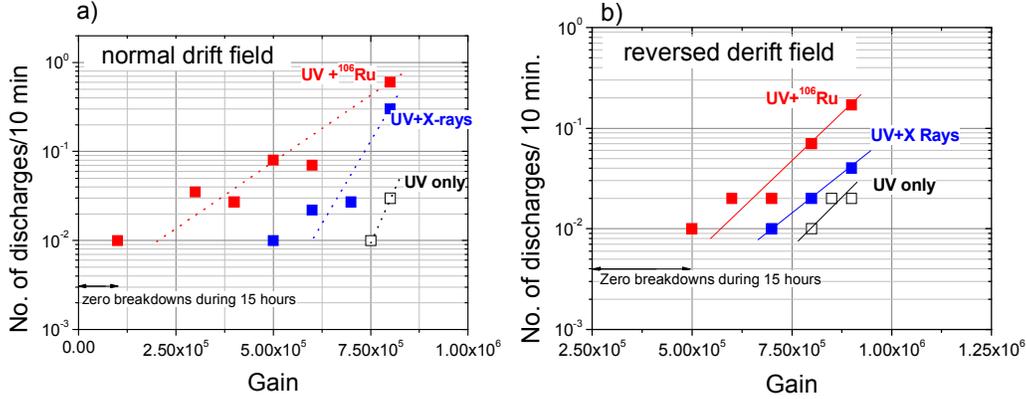

**Figure 8.** Breakdown rate of the CsI-triple-THGEM photon detector vs. gain, measured with UV photons and with UV plus low-rate $^{55}$Fe or $^{106}$Ru, in Ne+10%CH$_4$; a) under normal drift- field and b) under reversed drift field; $V_{drift}$ = 200V.

With UV irradiation only, the breakdown rate at normal drift-field polarization was rather low, of ~ 1 discharge per 6 hours, at the maximum gain ~ $8\times10^5$. In the presence of radioactive sources, particularly 106Ru, the breakdown rate at the same gain increased ~ 10-20 times. Over the whole investigated gain range, the sparking rate was considerably (~ 10 fold) lower with a reversed drift-field. It was practically independent of the UV-photon rate, suggesting its sole dependence on the radiation-induced avalanche-charge reaching the Raether limit.

The reversed-field approach seems to be effective in reducing discharge rate. Our measurements (carried out with a 6 cm$^2$ detector area and with irradiated area of ~ 1 cm$^2$) suggest that a triple-THGEM photon detector with reversed drift field, operating at gain of ~ $5\times10^5$, under similar UV and external radiation-background rates, will have 1-2 breakdowns per week per cm$^2$.

*Maximum gain and breakdown rates at high-intensity background*

The maximum achievable gain of the detector, or rather the maximum possible charge in the photon-induced avalanche, will naturally decrease with increasing ionizing background rate, due to sporadic overlap of the two. Systematic measurements were done with a CsI-triple-THGEM, operating in Ne+10%CH$_4$. It was simultaneously irradiated with the UV lamp ($N_{uv}$ ~ $10^5$ Hz/cm$^2$) and with X-rays from a Mo X-ray tube (mostly a Mo 18keV photo-peak and a weak broad-spectrum Bremsstrahlung), resulting in counting rate of $10^{2\text{-}}10^6$ Hz/mm$^2$. The maximum achievable gain, limited by x-ray induced sparks, is defined here as the one at which breakdowns appear at a rate of ~ 1 breakdowns per 10 min., is plotted in Fig. 9. The solid curve represents the general dependence of the maximum achievable total avalanche charge (primary ionization × gain) calculated in the same way as for the Cu tube (see above) vs. rate, valid for parallel-plate counters and for other micropattern gaseous detectors (see [13]). The maximum attainable charge of the CsI-triple-THGEM photon detector dropped with rate, similarly to the gain drop observed with a bare triple-THGEM (Fig. 6).

In other experiments (graphs not shown) with alpha-particles from $^{241}$Am (depositing ~$10^4$ electrons/cm in Ne/CH$_4$), a CsI-triple-THGEM reached a discharge limit at gains around $3\times10^3$ under normal drift field and at $2-5\times10^4$ (depending on the alpha-particle track orientation) under reversed drift field. We may conclude that in high-energy physics experiments, heavily ionizing particles' background may limit the maximum achievable gain of triple-THGEM photon detectors to values of few times $10^4$.

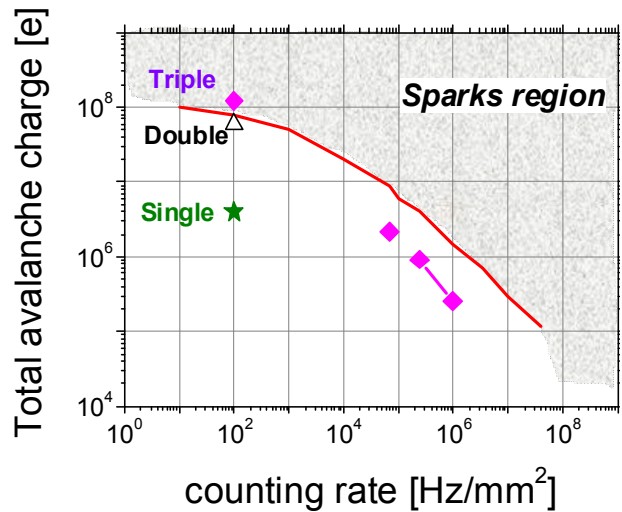

**Figure 9.** Maximum achievable total avalanche charge measured in single-, double- and triple-THGEMs with CsI photocathode (hole-diameter 0.4mm). The detector was irradiated simultaneously by a UV lamp and by X-rays from a Mo tube. Normal drift field; Ne+10%CH$_4$.

The main conclusion from the intense-background studies is that discharges in a CsI-coated triple THGEMs detector, irradiated simultaneously by UV photons and radioactive sources, were caused mainly by the source; this is in good agreement with Raether limit at low counting rates and with a previously-observed (in other gaseous detectors) general dependence of the maximum achievable avalanche charge on the rate at high counting rates. This established limit allows one to carefully assess applications of CsI-coated UV-photon THGEM detectors in RICH and in other applications, taking into consideration the anticipated background at the detector's vicinity.

*Cathode excitation effects*

As described in [13] and in some references therein, the intense ion bombardment of the CsI photocathode was observed to cause temporal increase of its sensitivity to UV and visible light. The CsI quantum efficiency (QE) resumed its original values only after 20-30 minutes, during which an increased rate of spurious pulses was observed. This phenomenon was named: "cathode excitation effect" ([13] and references therein).

It was further observed that the higher the avalanche-induced ion current density, the longer is the detector's recovery-time. The longest recovery times were observed after breakdown. It is assumed that this phenomenon depends, among others, on the ion species, namely on the gas mixture.

A frequent consequence of cathode excitation in CsI-MWPC detectors is that, following breakdown, the nominal operation voltage cannot be resumed for a substantial period of time [13]. This phenomenon was recently confirmed by the COMPASS RICH group [12], where occasional discharges led to very long recovery times (up to one day). To avoid these phenomena some of the COMPASS RICH CsI-MWPCs had to be operated at lower voltages, which naturally affected the photon detection efficiency [12].

The "cathode excitation effect" is explained by the temporal formation of a positive-ions layer on the CsI surface during ion bombardment, which creates intense local electric fields, reducing the work function and possibly inducing emission of electron jets (see more details in [13]). This phenomenon is directly related to the Malter effect [17].

We investigated the "cathode excitation effect" induction in a CsI- triple-THGEM, by intensely irradiating it with Mo X-rays. In a first series of measurements the photon detector, operating at a gain of $10^4$, was irradiated by a collimated X-ray beam, delivering $\sim 3\times10^4$ Hz/mm$^2$ photons to the irradiated spot. In these conditions the breakdown rate was $\sim 1$ per 10 min. After 5-10 minutes of irradiation, the X-ray tube was switched off and the rate of after-pulses, and pulses induced by a visible-light torch were measured (note that normally a CsI photocathode is "solar blind" with a cut-off at $\sim 220$ nm). The results are depicted in Fig. 10, showing a measurable rate of spurious pulses and visible-light pulses, lasting for about 30 minutes and fully subsiding after about 100 minutes. These results were reproducible, provided the detector was not irradiated for a couple of hours.

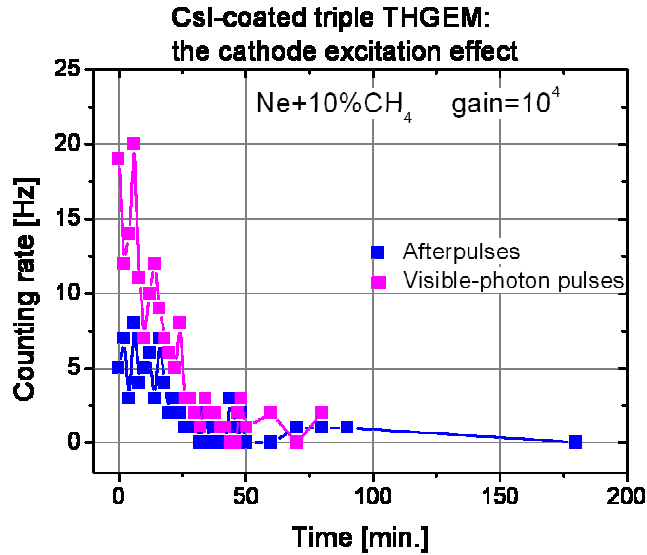

**Figure 10.** Counting rate of spurious pulses and visible-light induced pulses vs. time after induction of the cathode excitation effect; CsI-coated triple THGEM; gas mixture: Ne+10%CH$_4$; gas gain $\sim 10^4$.

In the following set of measurements, "cathode excitation" was induced in the same way but was followed by an immediate increase of the triple-THGEM voltages to a gain of $\sim 10^5$, ensuring sensitivity to single electrons. The results are presented in Fig. 11. Both spurious-pulse rate and visible-photon sensitivity increased by about a factor 10, but subsided after a period of time comparable to that of the lower-gain ($10^4$) operation.

The onset of discharges in these experiments always occurred in the last THGEM, at total gain of $2-3\times10^5$; the recovery time of the triple-THGEM, for resuming its voltages, was of about 30-60 sec. The data indicate that "cathode excitation" could be initiated in the CsI-triple-THGEM, with its characteristic increased spurious-pulses rate and sensitivity to visible light, but with no significant increase of recovery time.

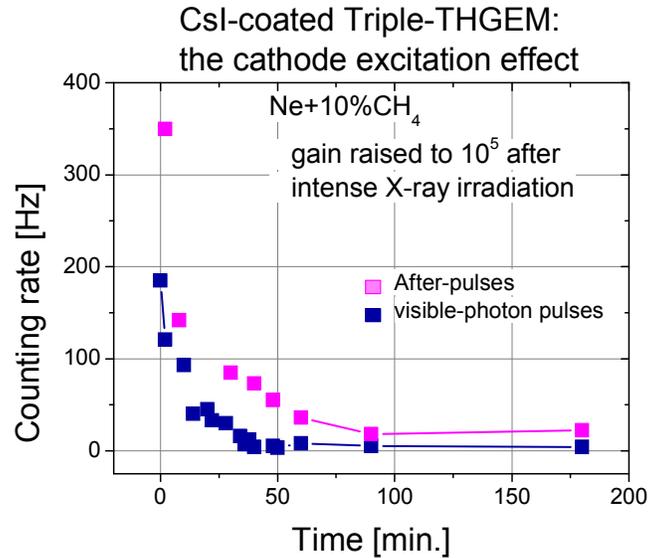

**Figure 11.** Counting rate of spurious pulses and visible-light pulses vs. time after cathode excitation induction at gain $10^4$, followed by a 10-fold gain increase. CsI-coated triple THGEM, Ne+10%CH$_4$ mixture, gas gain $\sim 10^5$.

The "cathode excitation effect" was also observed after intense UV irradiation; however in real experiments the RICH detector usually operates at much lower UV flux then that required for triggering a noticeable effect.

*Operation in pure CF$_4$ and in Ne/CF$_4$*

In some applications, it might be desirable replacing the detector-gas (e.g. pure CH$_4$) with a non-flammable one. Especially this is important for large-volume detectors (e.g. ALICE and COMPASS RICH). One of the options is CF$_4$. Recent studies [4, 18, 2, 19] indicated upon high photoelectron extraction efficiencies from CsI and their full collection into THGEM holes in pure CF$_4$ (similar to CH$_4$). This also makes CF$_4$ an attractive candidate for UV-detectors coupled to Cherenkov radiators through windows or, even more attractive, operating in a windowless mode with the same gas in both elements (e.g. like in the HBD described in [7] and discussed for ALICE-VHMPID [20]. Electron extraction and collection into detector holes in Ne/CF$_4$ mixtures showed also good performance, as discussed in [4].

Gains of ~$10^4$ were recorded in a single-THGEM in Ne/CF$_4$ (5%, 10%) under low-intensity X-ray irradiation (Fig. 5c-d). Under UV light irradiation, about 100-fold higher gains were reached in a single-THGEM, as depicted in Fig. 12. This reconfirms that even in highly scintillating gases emitting UV light, like CF$_4$ (see [21, 22] and references therein), CsI-THGEM breakdowns are initiated by the Raether limit and not by feedback processes.

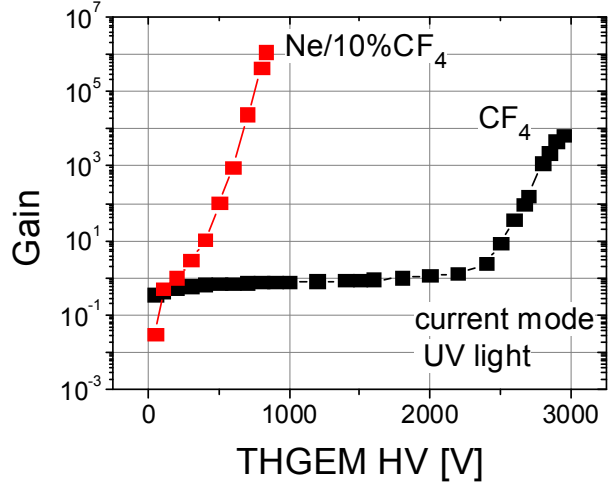

**Figure 12.** Gain measured in a singleCsI--THGEM (hole diameter 0.4mm) in current mode with UV photons; Ne+10%CF$_4$ and CF$_4$.

The gas gain obtained with triple-THGEM in Ne+10%CF$_4$, with X-rays at counting rates < $10^3$ Hz/mm$^2$, was ~ $10^5$ (Fig. 5); the same gain, adequate for RICH applications, was recorded when the detector was simultaneously irradiated with UV and high intensity of x-rays.

Figure 12 also shows the gain of a single CsI-THGEM in pure CF$_4$; the maximum gain under UV-irradiation was $10^4$, while in the presence of x-rays it was limited to ~ $10^2$. However, in the case of reversed drift field, the maximum achievable gain was close to $10^4$ even when the detector was simultaneously irradiated with UV and low intensity x-ray photons (reverse filed reduces the fraction of the collected primary ionization). The gas gain achieved in these preliminary studies is already as high as in The PHENIX triple-GEM HBD [23]. In contrast to [6], no irreversible damages of this THGEM were found in case of discharges, which is encouraging for possible applications. Preliminary measurements in CF$_4$ with a double-THGEM indicated upon higher reachable gains in pure CF$_4$; however, the higher operation voltages in this gas may be less convenient in practice, which deserves additional studies particularly for windowless Cherenkov radiator/detector systems.

*Gain-stability*

Several works investigated the THGEM's long-term gain stability under various experimental conditions [24, 25]. The main findings can be summarized as follows: in THGEMs having dielectric material exposed to avalanche electrons and ions, e.g. the rims around the holes and hole walls [1], the gas gain varies (increases or decreases) over time periods of up to several hours after voltage application. The gain variation is typically of a factor of two prior to stabilization. THGEMs without rims around the holes reached typically 10 fold lower gains [25, 1] compared to those with rims [3]; and exhibited much shorter stabilization times, of a fraction of an hour, almost independent of the counting rate. The existing data suggest that the long gain-stabilization time, also observed in GEMs [26], is related to charges slowly accumulating at the surface of the bare dielectric substrate (generally FR-4 or G-10) in the avalanche region. The time required for reaching equilibrium depends on the type of gas (positive ions), the electric fields, the avalanche charge and the rate, and on other parameters such as the amount of the exposed dielectric, the degree of adsorbed moisture (affecting surface and volume resistivity) etc. Note that the latter depend on detector operation conditions, e.g. detector pumping prior to gas introduction, time of gas flushing etc [24].

Most of the previously-published stability measurements were performed with THGEMs operating in Ar-based gases. First results in Ne-based mixtures, operating at considerably lower voltages [3], indicated upon a strong correlation between the stabilization time and the gas-flushing time prior to the measurement - suggesting a dependence on gas impurities (e.g. moisture).

The aim of the present measurements, of short- and long-term stability, respectively carried over hours and days, was to gain more understanding on this subject. In particular, we investigated stability at low counting rates (monitoring charge pulse heights), e.g. similar to those expected for the ALICE-VHMPID detector ($\leq 100$ Hz/cm$^2$), and compared the stability of bare and CsI-coated THGEMs. All THGEMs investigated had holes with 0.1mm rims.

*Short-term stability*

The results obtained with a single-THGEM confirmed those presented in [3]: ~ 60% gain increase in measurements starting immediately after gas introduction, compared to no gain increase in measurements done after 24 hours of gas flushing. Therefore, all the following measurements were done after gas-flushing for at least 24 hours in the detector.

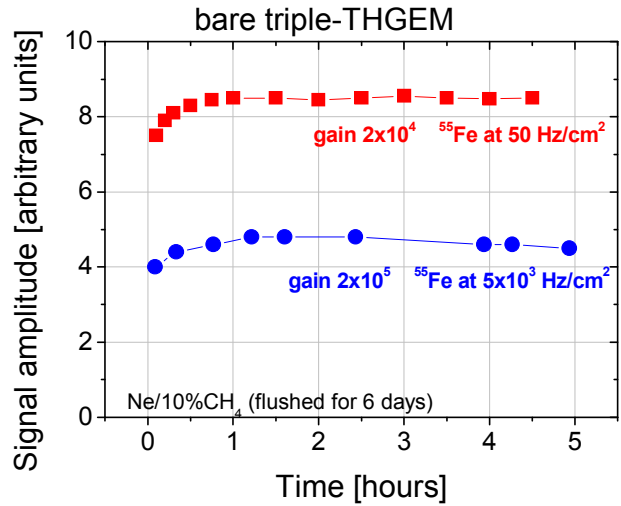

**Figure 13.** Short-term stability of a bare triple-THGEM (rim 0.1 mm), irradiated with $^{55}$Fe at gain $2\times10^4$, counting rate: ~ 50 Hz/cm$^2$ and at gain $10^5$, counting rate: ~ $5\times10^3$ Hz/cm$^2$. Gas mixture Ne+10%CH$_4$ flushed for 6 days prior to the measurement.

The stability results with a triple-THGEM, studied here for the first time, are presented in the following figures. The bare triple-THGEM was irradiated with a $^{55}$Fe source in two different conditions (Fig. 13): 1) counting rate of ~ 50 Hz/cm$^2$ and a gas gain of $1.8\times10^4$; 2) counting rate of 5 kHz/cm$^2$ and gas gain of $10^5$. At both conditions the bare triple-THGEM operated quite stably, with minor (< 15%) gain variations.

Figure 14 depicts the results for a single CsI-THGEM, with gain increase of ~25% during the first two hours. These results are in agreement with those presented in [3].

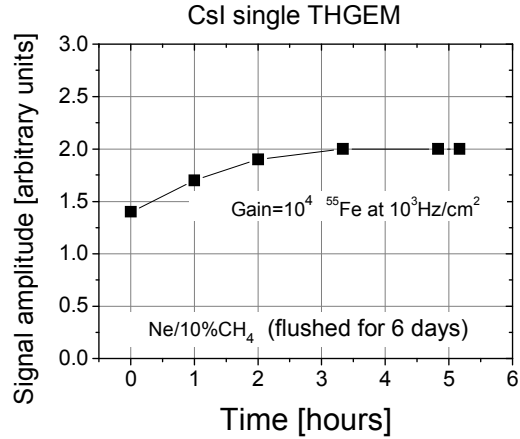

**Figure 14.** Short-term stability of a CsI-coated single-THGEM (rim 0.1 mm), operating in Ne+10%CH$_4$. Gas gain $10^4$, $^{55}$Fe counting rate ~1 kHz/cm$^2$.

The results obtained with a triple CsI-THGEM configuration of total gain ~$10^5$, at two different counting rates (1 and 5 kHz/cm$^2$) are presented in Fig. 15. Larger gain variations, up to 75%, were observed with the CsI-coated triple-THGEM.

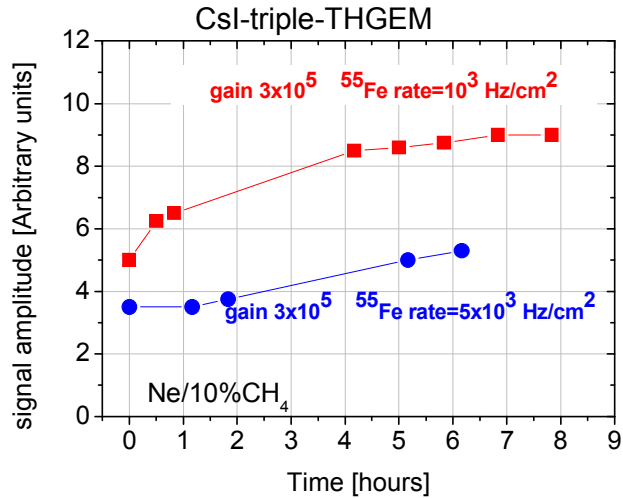

**Figure 15.** Short-term stability of a triple-THGEM (rim 0.1 mm) with CsI photocathode measured in Ne+10%CH$_4$ at a) at overall gain of $3\times10^5$ and counting rate of ~ 1 kHz/cm$^2$ and b) an overall gain of $3\times10^5$ and counting rate of ~ 5 kHz/cm$^2$.

The results of a short-term stability test of a CsI-coated single-THGEM operating in pure $CF_4$ at a gain of $3\times10^3$ are shown in Figure 16; one observes a gain increase of about 30% within 5-6 hours. Note the higher (>4 fold) THGEM operation voltage, ~ 3 kV in $CF_4$ compared to 0.7 kV in Ne/$CF_4$ for the same gain (see Fig. 12).

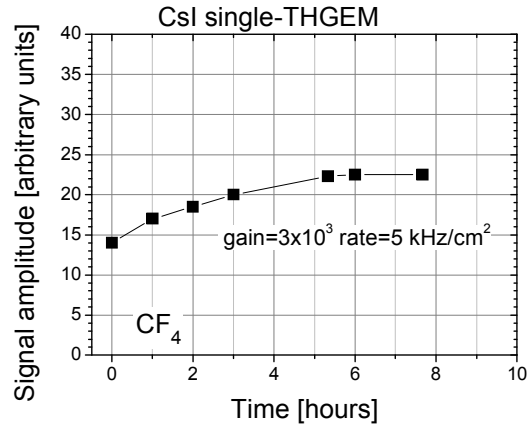

**Figure 16.** Short-term stability of a single-THGEM (rim 0.1mm) with CsI photocathode measured in $CF_4$ at a gain of $3\times10^3$ and counting rate of ~5kHz/cm$^2$.

*Long term stability*

The stability of bare and CsI-coated triple-THGEM was further investigated over a one-week period, in Ne/10%$CH_4$; the results presented in Fig. 17 are to be considered as preliminary. Following a fast rise, the gain of the bare triple-THGEM stabilized after two days; that of the CsI-coated THGEM kept rising during the entire week. In both cases we observed a total two-fold increase over a week. It is not possible at the moment to draw conclusions on the basis of these first results.

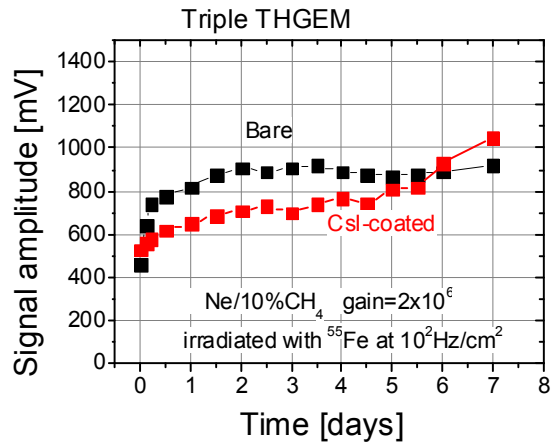

**Figure 17**. Long-term stability results of bare and CsI-coated triple-THGEM. Ne/10%$CH_4$; gain $2\times 10^6$; $^{55}$Fe x-rays, rate: $10^2$ Hz/cm$^2$.

## 4.2 Comparative studies with CsI-MWPC

A major aim of this work was the comparison between CsI-THGEM and presently used CsI-MWPC UV detectors. In [4] we investigated the UV-photon detection efficiency of CsI-coated THGEM detectors and compared it to the known efficiency of CsI-MWPC detectors used in COMPASS RICH. It was concluded that in $Ne/CH_4$ and $Ne/CF_4$ and with an optimized THGEM geometry (hole-diameter 0.3 mm, pitch 1 mm) the effective photon detection efficiency (combined quantum efficiency, extraction efficiency and collection efficiency) would reach values of 0.20-0.23 at 170 nm. This value is close to that of the photon detection efficiency currently achieved with the COMPASS MWPC, where it is primarily limited by the low gain ($\sim 10^4$) in beam conditions, resulting in poor signal-over-threshold. We have speculated that with a cascaded-THGEM detector, a 10-fold higher gain is expected, with a larger attainable signal-over-threshold value.

In this work we focused on the maximum achievable gain, rate characteristics, feedback processes and "cathode excitation effects". The studies with MWPCs were conducted with a bare stainless steel cathode (bare-MWPC) and with a cathode coated with CsI (CsI-MWPC). The MWPC was flushed with $CH_4$ (used in ALICE and COMPASS), $Ne+10\%CH_4$ or pure $CF_4$ at 1 atm.

*Gas gain and stability*

Figure 18 depicts the gain curves of the bare- and CsI-MWPCs, measured in $CH_4$ with a $^{55}Fe$ source. At gains below $3\times10^4$ the signals were recorded with a charge sensitive amplifier (Ortec 142PC); at higher gains, the pulses were recorded with a fast preamplifier VT110CH4 (ESN electronics).

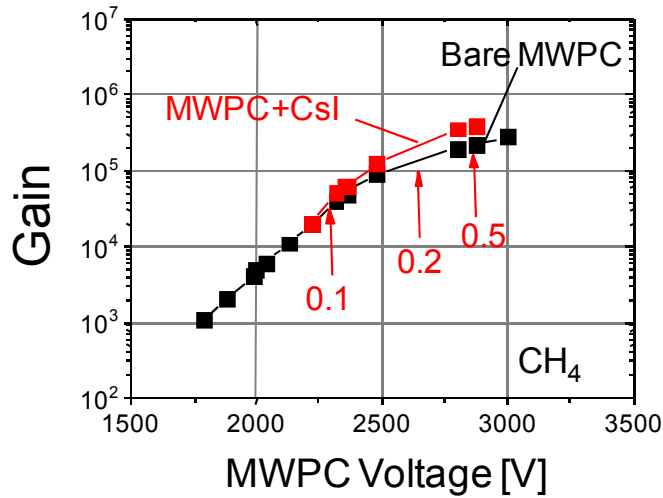

**Figure 18.** Gain curves of the bare MWPC and the CsI-MWPC, measured with UV light (for gains below $10^4$) and with $^{55}Fe$ x-rays (for gains above $10^3$ The arrows indicate the gains at which feedback pulses appeared in the CsI-MWPC and their amplitude ratio η; gas: $CH_4$.

As seen in Fig. 18, for voltages below 2250 V (gain $< 5\times10^4$), the gas gain increased exponentially with the voltage; gain-saturation due to space charge, typical to wire chambers, started at voltages above 2500V (gain $> 10^5$).

The gain curves for bare- and CsI-MWPC at voltages below 2500 V are almost identical, however, in CsI-MWPC after-pulses (secondary avalanches) appeared already at gains $> 3\times10^4$, and they were delayed by about 40ns with respect to the primary pulses. The arrows in Fig. 18 indicate the gains at which feedback pulses appeared in the CsI-MWPC and the numbers correspond to the amplitude ratio η of the secondary-to-primary pulses: η = A2/A1; η increased with gain, from η = 0.1 at $3\times10^4$, to 0.2 at $2\times10^5$. The probability of photon feedback $\gamma_{fp}$ (Eq. 4) can also be written as (see [13]):

$$\gamma_{fp} = \int Q(V, E_v)\, S(V, E_v)\, dE_v \qquad (5)$$

where $Q(V, E_v)$ is the quantum efficiency of the CsI photocathode in gas as a function of the voltage V applied to the detector and the photon energy $E_v$, and $S(V, E_v)$ is the photon emission spectrum of the Townsend avalanches at voltage V [13]. We have marked that $\gamma_{fp}$ (or equivalently $G_f$) in our CsI-MWPC and in ALICE- and COMPASS-RICH detectors are very similar. Using Eq. 5 and assuming the emission spectra $S(V, E_v)$ in the present CsI-MWPC and in that of ALICE- and COMPASS-RICH detectors are the same, one can conclude that the quantum-efficiency function $Q(V, E_v)$ of the present (not measured directly) and the other detectors (with well know quantum efficiency) must be quite similar. Therefore the present CsI-MWPC is a realistic representative of the other detectors for a valid comparison with the CsI-THGEM.

At voltages above 3200 V (gain $\sim 10^6$) a corona discharge appeared in CsI- MWPC due to the feedback mechanism: $G_{f\gamma f} = 1$. Thus, in contrast to the CsI-triple-THGEM, in which the breakdown was triggered mostly by the x-rays or beta-particles reaching the Raether limit, in the CsI-MWPC these particles did not induce breakdowns at gains below the critical one. This is a fundamental difference between both photon detectors: the discharge onset in the CsI-MWPC seems to be solely due to photon-feedback.

In the bare MWPC no feedback pulses were observed at $V_{MWPC} < 3500$ V; the onset of corona discharge was observed at voltages around V ~ 3800.

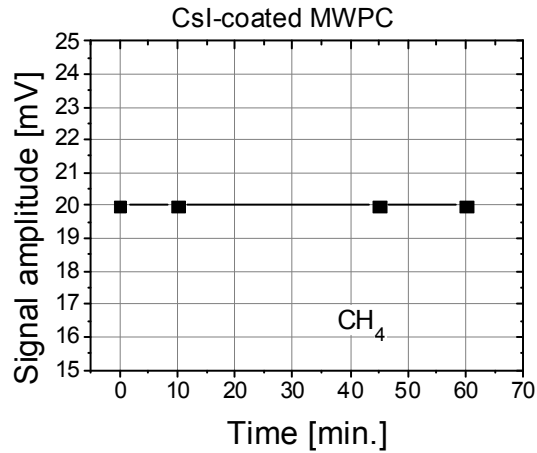

**Figure 19.** A short-term stability of the CsI-MWPC, irradiated with $^{55}$Fe x-rays at gain $10^4$ and at a rate of 5 kHz/cm$^2$. Gas: CH$_4$.

Figure 19 shows the results of the short-term stability of the CsI-MWPC operated with $^{55}$Fe x-rays at gain $10^4$ and at a rate of 5 kHz/cm$^2$, in CH$_4$. Long-term stability measurements at a similar gain, but at considerably lower counting rates ($< 10^3$ Hz/cm$^2$), were performed by ALICE group [27]; they were remarkably good, yielding ~ 30% gain variations over a time period of more than one year.

*Operation in CF$_4$*

CF$_4$ is currently considered by ALICE VHMPID group as an alternative to CH$_4$. It would permit the use of a windowless radiator/detector configuration; being inflammable it would satisfy safety requirements. Figure 20 presents gain curves of the CsI-MWPC measured in Ne+10%CF$_4$ and in pure CF$_4$. The operation voltages of the MWPC in CF$_4$ were similar to those in CH$_4$, and, like in CH$_4$, at gains above $10^4$ feedback pulses appeared, noticeable as gain increase, deviating from the exponential curve. These preliminary measurements indicate that CF$_4$ filled CsI-MWPC can be an attractive option for the ALICE RICH Additional measurements in CsI are required to demonstrate long-term stability in CF$_4$.

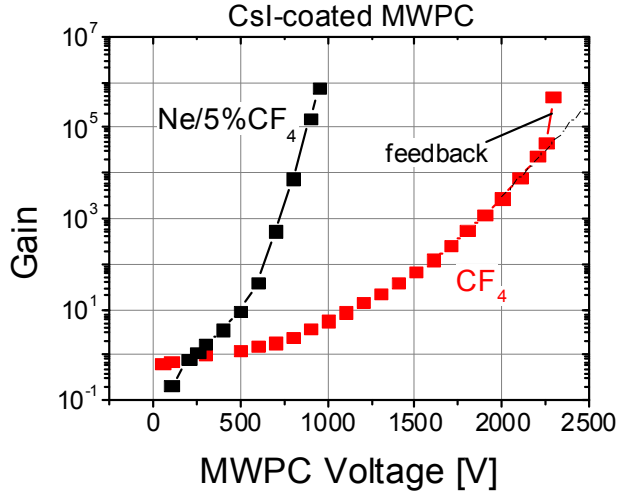

**Figure 20.** Gain curves of the CsI-MWPC measured in Ne+5%CF$_4$ and in pure CF$_4$, with UV photons

*Cathode excitation effects*

Similarly to triple-THGEM, we investigated the "cathode excitation effect" in the CsI-MWPC, induced by intense irradiation with a Mo X-ray gun. The x-ray intensity was kept similar to that used in the experiments with triple-THGEMs (see Fig. 10, 11). Figure 21 depicts the counting rate vs. time of spurious after-pulses and pulses induced by visible light, following the exciting x-ray irradiation. An elevated rate of both types of pulses was observed for more than three hours; after 4-5 hours it resumed its original level of ~ 2 Hz and ~ 20 Hz for spurious pulses and visible-photon pulses, respectively. The cathode excitation effect at a gain of $5\times10^4$ was by far more intense compared to the triple-THGEM (investigated at gains of $10^4$ and $10^5$, see Fig. 10, 11 respectively), manifesting both higher rate of spurious pulses and elevated sensitivity to visible light for longer periods of time. This may be due to the "open geometry" of the MWPC in which the CsI photocathode is receptive to the full avalanche-ion and -photon flux. In contrast, at the same gain, only a small fraction of the avalanche ions will reach the photocathode in the cascaded-THGEM (for double-THGEM see [8]), similarly to cascaded-GEM [28]. The ion backflow in cascaded-THGEM can be further reduced, e.g. with ion-trapping strips patterned on the hole-electrodes [29], as demonstrated with Microhole & Strip Plate (MHSP) hole-multipliers [30].

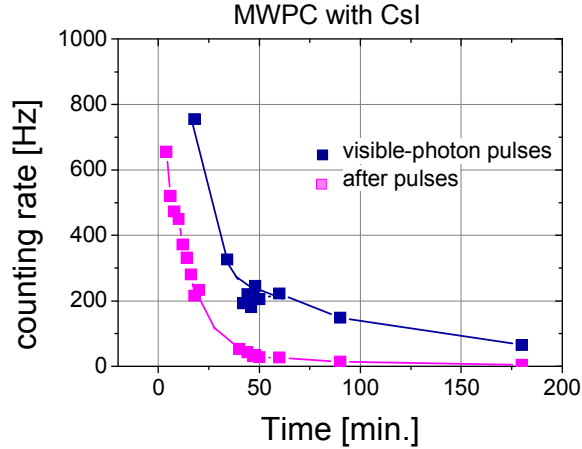

**Figure 21.** Counting rate vs. time of spurious pulses and pulses induced by visible photons, measured after the CsI-MWPC irradiation with an intense x-ray beam. The x-ray radiation was stopped at time zero. Gas: $CH_4$; gain: $5\times10^4$.

In a following measurement we applied voltage on the MWPC, sufficient to trigger a corona discharge. Following this discharge, and even though it lasted for only a few seconds, we could no more apply the voltage for almost one hour. This is different from similar experiments with THGEM, in which high voltage could be restored in ~ 30 sec following a discharge. Similar long recovery times of CsI-MWPC were observed in COMPASS RICH [12]. This difference may be similarly attributed to the fact that THGEMs screen the majority of discharge-ions and photons.

As mentioned above, discharges in the MWPC appeared only at some critical gain; at lower gains, no discharges were triggered by low- or medium-intensity sources. However, we observed that at very high x-ray intensities the value of the critical gain was reduced by about a factor of two.

The main conclusion drown from these studies is that the cathode excitation effect is more intense in CsI-MWPC than in CsI-triple-THGEM. It should be seriously taken into consideration in the choice of high-rate UV detectors for RICH applications.

## 5. Summary and discussion

In this work we presented new results on the operation of single-, double- and triple-THGEM detectors, in $Ne/CH_4$ and $Ne/CF_4$ mixtures. The work focused on UV-detectors with reflective CsI photocathodes deposited on the first THGEM electrode, in view of their potential applications in RICH systems. The studies were carried out with soft x-rays and UV-photons. In some experiments with UV-photons the detectors were simultaneously irradiated with x-rays or β-electrons, simulating high-ionization background. The research encompassed studies of the maximum achievable gain at various conditions, discharge probability, cathode excitation effects and gain stability. Comparative studies were made in similar experimental conditions with a MWPC coupled to a reflective CsI photocathode, of a geometry similar to that used in CERN-ALICE and -COMPASS RICH detectors; the detector gases were $CH_4$ (RICH-standard), $CF_4$ and $Ne/CF_4$.

Systematic studies of gain limits due to rate-induced discharges were performed with soft x-rays for various bare- and CsI-THGEM configurations (single, double and triple) at radiation flux between $10^3$ and $10^6$ $Hz/mm^2$. It was shown that at very high flux, the gain was significantly reduced. However, independently of the exact gain-drop mechanism with rate, the measured maximum achievable gain of a single- and cascaded-THGEM (Figures 6 and 9) followed the general trend observed in most gas-avalanche detectors [13]; it is within a "permissible-zone", depending on primary ionization, gain and rate.

In some gaseous photon detectors incorporating a photocathode, avalanche-induced feedback mechanisms may trigger breakdowns; e.g. ion-induced gain limits in GEM-based visible-sensitive detectors with bialkali photocathodes [15]; these were dramatically reduced with cascaded hole-multipliers having ion-trapping strips [31]. In CsI-THGEMs, feedback effects are significantly reduced and gain limitations were not observed. Moreover, it was clearly demonstrated that the maximum gain, at low counting rates, is governed by the Raether limit. As a consequence, a triple CsI-THGEM operated stably at gains ~ $3-5\times10^5$ in the presence of low-intensity MIPs radioactive background (see Fig. 7, 8); in the presence of a higher-ionizing x-ray background, the maximum achievable gain is lower, limited by the x-ray induced total charge (see Fig. 9). The gain will further drop in the presence of heavily ionizing particles. With reversed drift field one can reach much higher gains, thus strongly competing with MWPCs. Thus, when considering THGEM for RICH, one should carefully assess the nature and intensity of the ionization background and correctly choose the mode of operation.

Another problem addressed here was the THGEM's gain-stability over time. It was already established in the literature that THGEMs without etched rims around the holes have stable gain on the short-range (hours) and the long-range (weeks) time scales [10]. However, the lack of rims enhanced discharge probability due to defects on the hole-edges, resulting in over 10-fold gain reduction compared to holes with etched rims [1]. Holes with rims showed gain variations in amplitude over time; these depended on the substrate material, gas and impurities (e.g. moisture affects surface and volume resistivity), hole-geometry, applied voltage etc. It was found for example that the operation in Ne-mixtures (at lower voltages) resulted in better gain stability, even with 0.1 mm rims [3]. The results presented here with bare THGEM electrodes and others coated with CsI, with 0.1 mm rims around holes, are not yet conclusive. Gain variations up to a factor of two were observed in different conditions. Though not dramatic for detector operation, more systematic studies are required on rim-size optimization.

A considerable advantage of a CsI-triple-THGEM is the fact that, due to the Raether limit, discharges occur mostly in the last THGEM element (final avalanche). Since only a fraction of the avalanche ions created in the discharge reach the photocathode, cathode excitation effects are not promoted and thus triple-THGEM detectors have short recovery time (about 30 sec) following an eventual discharge. We have also shown that in a CsI-triple-THGEM, with the CsI-photocathode relatively screened from avalanche ions, the rate of spurious electron emission due to cathode-excitation effects is considerably reduced compared to that of a CsI-MWPC.

Because discharges in CsI-MWPC occur primarily via a feedback loop ($G_{f\gamma f} = 1$), if $\gamma_f$ increases under intense ion bombardment (the "cathode excitation" effect), than the conditions for discharge are fulfilled at lower gas gains. This could be the reason for stability loss in CsI-MPWC at high counting rates, with transition to discharge - as was observed for example in COPASS RICH [12].

Our studies accentuated two main differences between a CsI-THGEM and CsI-MWPC:
1) Due to its open geometry, feedback pulses in a CsI-MWPC appear already at rather low gas gains, limiting its operation stability; e.g. in $CH_4$ 10% avalanche-feedback appears already at gain of $5\times10^4$. We did not observe feedback effects in CsI-THGEM even at considerably higher gains.
2) Corona discharge in MWPC appears at a certain maximal gain $G_m=1/\gamma$, practically independent on the nature of the ionizing radiation or the counting rate.

As a consequence, at low counting-rates the CsI-MWPC operated at a gain almost ten-fold lower than the CsI-THGEM, although no discharges appeared even with heavily ionizing particles. At high counting-rates, at a fixed voltage, the CsI-MWPC gain dropped due to space charge, e.g., by a factor of 3 (from $2\times10^4$ to $7\times10^3$ (see Fig. 22) when increasing the rate to $10^5$ Hz/mm$^2$. This is comparable to a gain at which a CsI-THGEM with a normal drift field can operate stably at the same counting rate (see Figure 9).

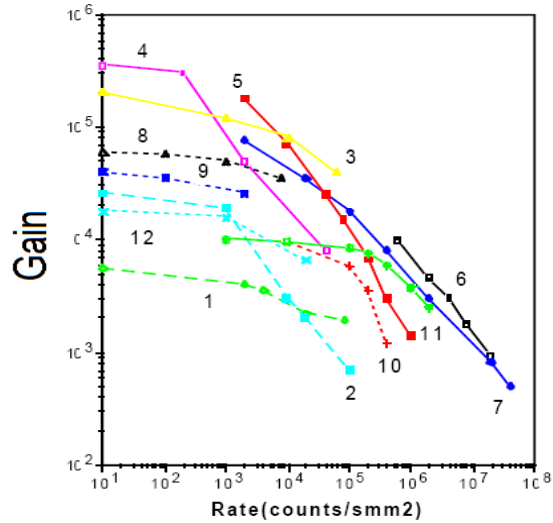

**Figure 25.** The maximum achievable gain (curves 1-7), as a function of X-ray flux, for various detectors: 1) diamond-coated MSGC (1 mm pitch); 2) diamond-coated MSGC (1-mm pitch); 3) MSGC (2 mm pitch) combined with GEM; 4) MSGC (1 mm pitch) combined with GEM; 5) PPAC (3 mm gap); 6) MICROMEGAS; 7) thin gap (0.6 mm) PPAC. Curves 8-12 show the space-charge gain limit vs. rate in MWPC at different gains (from ref. [32]).

To summarize:
1. A CsI-THGEM can operate without feedback at gas gains higher than a CsI-MWPC, but only in the absence of elevated ionizing background. Its maximum achievable gain is governed by the Raether limit.
2. With reversed drift-filed, CsI-THGEMs can operate at high gains even in the presence of ionizing particles.

3. A CsI-MWPC is more immune to heavily ionizing particles, which do not cause discharges; the latter are rather due to feedback processes.
4. At high courting rates CsI-MWPC is more prone to instabilities than CsI –THGEMs due to the higher probability to develop cathode excitation effects.

The present work provided additional information on THGEM operation, particularly on UV detectors with CsI photocathodes. With expected photon detection efficiency close to CsI-MWPC (currently investigated in beam experiments), CsI-THGEM UV-detectors could become suitable candidates for high-rate single-photon imaging. The choice of UV-detector for RICH will naturally depend on the application counting rates and on the nature and rate of the ionizing background in a given experiment.

## Acknowledgments

This work was partially supported by the Israel Science Foundation, grant Nº 402/05 and by the Benoziyo Centre for High Energy Physics Research. It was pursued within the framework of the CERN RD51 collaboration. V.P. acknowledges the Weizmann Institute for his Visiting-Professor Fellowship. We thank M. Klin for his technical support and Oren Cohen for his contribution to some of the measurements. A. Breskin is the W.P. Reuther Professor of Research in the peaceful use of Atomic Energy.

## References

[1] A. Breskin, R. Alon, M. Cortesi, R. Chechik, J. Miyamoto, V. Dangendorf, J.M. Maia, J.M.F. Dos Santos; *A concise review on THGEM detectors*. NIMA **598** (2009) 107-111.

[2] R. Chechik, A. Breskin; *Advances in gaseous photomultipliers*. NIMA **595** (2007) 116-127.

[3] M. Cortesi, V. Peskov, G Bartesaghi, J. Miyamoto, S. Cohen, R. Chechik, J.M.F. dos Santos, J.M. Maia, G. Gambarini, V. Dangendorf and A. Breskin; *THGEM operation in Ne and Ne/CH$_4$*. 2009 *JINST* **4** P08001.

[4] C.D.R. Azevedo, M. Cortesi, A.V. Lyashenko, A. Breskin, R. Chechik, J. Miyamoto, V. Peskov, J. Escada, J.F.C.A. Veloso and J.M.F. dos Santos; *Towards THGEM UV-photon detectors for RICH: on single-photon detection efficiency in Ne/CH$_4$ and Ne/CF$_4$*. 2010 JINST **5** P01002.

[5] R. Alon, J. Miyamoto, M. Cortesi, I. Carne, A. Breskin, R. Chechik, J. M. Maia, J.M.F. dos Santos, M. Gai and D. McKinsey; *Operation of a Thick Gas Electron Multiplier (THGEM) in Ar, Xe and Ar-Xe. 2008 JINST* **3** P01005.

[6] R. Chechik, A. Breskin and C. Shalem; *Thick GEM-like multipliers - a simple solution for large area UV-RICH detectors*. NIMA **553** (2005) 35-40.

[7] Z. Fraenkel, A. Kozlov, M. Naglis, I. Ravinovich, L. Shekhtman, I. Tserruya, B. Azmoun, C. Woody, S. Sawada, S. Yokkaichi, A. Milov, T. Gunji, H. Hamagaki, M. Inuzuka, T. Isobe, Y. Morino, S.X. Oda, K. Ozawa, S. Saito, T. Sakaguchi, et al. *A hadron blind detector for the PHENIX experiment at RHIC. NIMA* **546** (2005) 466-480.

[8] C. Shalem, R. Chechik, A. Breskin, K. Michaeli; *Advances in Thick GEM-like gaseous electron multipliers—Part I: atmospheric pressure operation*. NIMA **558** (2006) 475-489.

[9] V. Peskov; *R&D on CsI-TGEM based photodetector* presented at the 7th International Workshop on Ring Imaging Cherenkov detectors (RICH 2010), 2-7 May 2010; Cassis, France.; Available online at: http://indico.in2p3.fr/contributionDisplay.py?contribId=20&sessionId=15&confId=1697

[10] M. Alexeev, M. Alfonsi, R. Birsa, F. Bradamante, et al.; *Development of THGEM-based photon detectors for Cherenkov Imaging Counters.* 2010 JINST **5** P03009.

[11] E. Nappi; *Trends in the development of large area photon detectors for Cherenkov light imaging applications*. NIMA **504** (2003) 70-87.

[12] S. Dalla Torre; *Status and perspectives of gaseous photon detectors.* Available online at: http://indico.in2p3.fr/contributionDisplay.py?contribId=102&confId=1697. F. Tessarotto, *The experience of building and operating COMPASS RICH-1.* Available online at: http://indico.in2p3.fr/contributionDisplay.py?contribId=35&sessionId=37&confId=1697. presented at the 7th International Workshop on Ring Imaging Cherenkov detectors (RICH 2010), 2-7 May 2010; Cassis, France.

[13] V. Peskov, P. Fonte. *Research on discharges in micropattern and small gap gaseous detectors*; arXiv:0911.0463. 2009

[14] H. Raether, *Electron avalanches and breakdown in gases*; Butterworths, London, 1964.

[15] D. Mörmann, M. Balcerzyk, A. Breskin, R. Chechik, B.K. Singh and A. Buzulutskov. *GEM-based gaseous photomultipliers for UV and visible photon imaging*. NIMA **504** (2003) 93.

[16] A. Lyashenko, A. Breskin, R. Chechik; *Ion-induced secondary electron emission from K-Cs-Sb, Na-K-Sb and Cs-Sb photocathodes and its relevance to the operation of gaseous avalanche photomultipliers. J. Appl. Phys.* **106** (2009) 044902.

[17] L. Malter, Anomalous Secondary Emission, a new phenomenon. *Phys. Rev.* **49** (1936) 478.

[18] J. Escada, L.C.C. Coelho, T.H.V.T. Dias, J.A.M. Lopes, J.M.F. dos Santos and A. Breskin; *Measurements of photoelectron extraction efficiency from CsI into mixtures of Ne with $CH_4$, $CF_4$, $CO_2$ and $N_2$. 2009 JINST* **4** P11025.

[19] A. Breskin, A. Buzulutskov, R. Chechik, *GEM photomultiplier operation in $CF_4$*. NIMA **483** (2002) 670–675.

[20] A. Agocs, R. Alfaro, G.G. Barnafoldi, et al.; *Very high momentum particle identification in ALICE at the LHC.* Nucl. Instrum. Meth. A 617 (2010) 424-429.

[21] M.M.F.R. Fraga, F.A.F. Fraga, S.T.G. Fetal, L.M.S. Margato, R. Ferreira Marques, A.J.P.L. Policarpo; *The GEM scintillation in He–$CF_4$, Ar–$CF_4$, Ar–TEA and Xe–TEA mixtures*. NIMA **504** (2003) 88-92.

[22] A. Morozov, M.M.F.R. Fraga, L. Pereira, L.M.S. Margato, S.T.G. Fetal, B. Guerard, G. Manzin, F.A.F. Fraga; *Photon yield for ultraviolet and visible emission from $CF_4$ excited with α-particles*. NIMA **268** (2010) 1456-1459.

[23] I. Tserruya et al.; *Design, Construction, Operation and Performance of the Hadron Blind Detector for the PHENIX Experiment at RHIC*. Presented at RICH10, April 2010, Cassis, France. http://indico.in2p3.fr/contributionDisplay.py?contribId=14&sessionId=37&confId=1697

[24] R. Chechik, M. Cortesi, A. Breskin, D. Vartsky, D. Bar, V.; *Thick GEM-like (THGEM) detectors and their possible applications.* Physics/0606162, 2006.

[25] M. Alexeev, R. Birsa, F. Bradamante, et al.; *The quest for a third generation of gaseous photon detectors for Cherenkov imaging counters. NIMA* **610** (2009) 174-177.

[26] B. Azmoun, W. Anderson, D. Crary, et al.; *A Study of Gain Stability and Charging Effects in GEM Foils. IEEE Nucl. Sci. Sym. Conf. Rec.* **6** (2006*);* DOI: 10.1109/NSSMIC.2006.353830.

[27] P. Martinengo, private communication.

[28] D. Mörmann, A. Breskin, R. Chechik, D. Bloch, *Evaluation and reduction of ion back-flow in multi-GEM detectors. NIMA* **516** (2004) 315.

[29] J. Veloso et al. *Thick-COBRA, a New Thick-Hole Concept for Ion Back Flow Reduction.* Presented at RICH10, April 2010, Cassis, France.

[30] A. Lyashenko, A. Breskin, R. Chechik, J. M. F. Dos Santos, F. D. Amaro, J. F. C. A. Veloso; *Efficient ion blocking in gaseous detectors and its application to gas-avalanche photomultipliers sensitive in the visible-light range. NIMA* **598** (2009) 116-120

[31] A. Lyashenko, A. Breskin, R. Chechik, F. D. Amaro, J. Veloso and J.M.F. Dos Santos; *Gaseous Photomultipliers for the visible spectral range. 2009 JINST* **4** P07005 and references therein.

[32] P. Fonte, V. Peskov, B.D. Ramsey; *Which Gaseous Detector is the Best at High Rates?*, Summer 1998 Issue, SLAC-JOURNAL-ICFA-16.